\documentclass[published]{JHEP3} 

\JHEP{00(2007)000}

\JHEPspecialurl{http://jhep.sissa.it/JOURNAL/JHEP3.tar.gz}

\usepackage{epsfig,multicol,bbm}

\newcommand\fverb{\setbox\fverbbox=\hbox\bgroup\verb}
\newcommand\fverbdo{\egroup\medskip\noindent%
            \fbox{\unhbox\fverbbox}\ }
\newcommand\fverbit{\egroup\item[\fbox{\unhbox\fverbbox}]}
\newbox\fverbbox

\title{Geometry of Dark Energy\\{\small(The  dynamics of the extrinsic curvature)}}

\author{M. D. Maia, A. J. S. Capistrano \\
    Universidade de Bras\'{\i}lia, Instituto de F\'{\i}sica, Bras\'{\i}lia - DF 70919-970, Brasil\\
    E-mail: \email{maia@unb.br}; E-mail: \email{capistranoaj@unb.br}}

\author{J. S. Alcaniz \\
    Observat\'orio Nacional, 20921-400, Rio de Janeiro - RJ, Brasil\\
    E-mail: \email{alcaniz@on.br}}

\author{E. M.  Monte \\
    Departamento de F\'{\i}sica, Universidade Federal da Para\'{\i}ba, 58059-970, Jo\~ao Pessoa - PB, Brasil\\
    E-mail: \email{edmundo@fisica.ufpb.br}}

\received{}       
\accepted{}     

\preprint{\hepth{9912999}}  

\abstract{The  acceleration of the universe is  described  as  a  dynamical effect  of the  extrinsic  curvature of  space-time.
By extending previous results,  the  extrinsic  curvature is  regarded   as  an  independent  spin-2  field, determined by a set of  non-linear equations similar to Einstein's  equations. In this framework, we  investigate some cosmological consequences of this class of scenarios and test its observational viability by performing a statistical analysis with current type Ia Supernova data.}

\keywords{Cosmology, Extra Dimensions, Dark energy}

\newcommand{\be}{\begin{equation}}
\newcommand{\ee}{\end{equation}}
\newcommand{\rf}[1]{(\ref{eq:#1})}

\begin{document}


\section{Introduction}

Modifications of gravity at very large scales constitute an alternative route to deal with the accelerated  expansion of the universe, often described  by  something called dark energy. That route in turn has  been predominantly associated  with  the existence of extra  dimensions, an idea that has been  explored   in various  theories beyond the standard model of particle physics, especially in theories for unifying gravity and the other fundamental forces, such as superstring or M theories. As suggested in \cite{CarterUzan}, extra dimensions may also provide a possible explanation for the huge difference between the two fundamental energy scales in nature, namely, the electroweak and Planck scales [$M_{Pl}/m_{EW} \sim 10^{16}$]. An important contribution to  this idea was subsequently given by  Arkani-Hamed {\it{et al.}} \cite{ADD}  inaugurating the  brane-world program and showing   that if our  world is embedded in a higher dimensional space, then  the  gravitational field  can  propagate  in the  extra dimensions,  keeping   ordinary matter  and  gauge  fields  confined  to our four-dimensional submanifold. The impact of  such program  in theoretical and observational cosmology  has  been discussed  at  length as,  e.g.,  in Refs.~\cite{RS,RS1,DGP,BWcosmology,BW1,BW2,BW3,BW4,BW5,BW6,BW7,BW8,BW9}. However, these are mostly  based on  specific  models  using special  conditions. For such large scale phenomenology   as the   expansion of the universe,   a  general theory based on fundamental  principles and  on solid   mathematical foundations is  still lacking.

In a previous communication \cite{GDEI} (hereafter referred to as paper I) we have studied possible modifications imposed on the Friedmann equation,  when the standard cosmological model is regarded as an embedded   space-time, within the covariant (model independent) formulation of the brane-world  program. By comparing the contribution  of the extrinsic curvature to the gravitational equations with a phenomenological quintessence model with a constant equation of state (EoS) $w$, we have found that the observed acceleration of the universe can be explained as an effect of the extrinsic curvature. However, the comparison  of  geometry with   a phenomenological fluid as  in paper I should be  seen   only as a  temporary  analogy, which was  necessary  because most  of the  known  phenomenology of the  accelerated  expansion of the universe is  based on   fluid mechanics.
Further  studies  on  the  theory of  embedded  Riemannian geometries have  shown  that  the extrinsic  curvature  acts  as  an independent  field   which  generates  the  extra dimensional  kinematics  of the  gravitational  field.

The goal of the present paper is twofold. First, to present a  mathematically  correct  structure  of   space-time embedding  based on Nash's  theorem.  Specifically, we  show  that in spite  of the fact  that  space-times are four-dimensional,   the gravitational  field   necessarily  propagates along the extra  dimensions of the embedding space. Second, to introduce the extrinsic  curvature
as  an independent   spin-2   field, described  by  an Einstein-like equation derived  by  S. Gupta. The paper  is  organized  as  follows: In  the following section  we give a brief  review  of paper I where  the extrinsic  curvature was  compared  with  $w$-fluid.  In  Sec.  III,  we  abandon   the  fluid  analogy, describing   the extrinsic  curvature as a dynamical  field derived  from Nash's  theorem and  on Gupta's  equations  for  a  spin-2  field in an embedded space-time. To test the viability of this scenario, we investigate constraints on the model parameters from distance measurements of type Ia supernovae (SNe Ia) in Sec. IV. We end this paper by summarizing our main results in Sec. V.

\section{The Geometric  Fluid Model}

In paper I, the  Friedmann-Lema$\hat{\mbox{i}}$tre-Robertson-Walker (FLRW) line element was  embedded in a  5-dimensional space with  constant curvature bulk space  whose  geometry  satisfy  Einstein's  equations with a cosmological constant
\be
{\cal R}_{AB} -\frac{1}{2} {\cal R} {\cal G}_{AB}  + \Lambda_* {\cal G}_{AB}=\alpha_*  T^*_{AB} \label{eq:BE0}\;,
\ee
where $T^*_{AB}$ denote the energy-momentum tensor components of the known material sources,  essentially  the
cosmic fluid, composed of ordinary matter interacting with gauge fields, confined  to the 4-dimensional  space-time.

When the above equations are written in the Gaussian frame defined by the embedded space-time,   we obtain a  larger   set of gravitational field equations
\begin{eqnarray}\label{eq:BE1}
&&R_{\mu\nu}-\frac{1}{2}Rg_{\mu\nu}+\Lambda_* g_{\mu\nu}-Q_{\mu\nu}=
-8\pi G T_{\mu\nu}\; \hspace{2mm}\\
&&\label{eq:BE2} k_{\mu;\rho}^{\;\rho}-h_{,\mu} =0\;,\hspace{4,9cm}
\end{eqnarray}
where  $k_{\mu\nu}$ denotes the extrinsic  curvature and
the quantity $Q_{\mu\nu}$  is a purely geometrical term given by
\begin{equation}\label{eq:qmunu}
Q_{\mu\nu}=g^{\rho\sigma}k_{\mu\rho }k_{\nu\sigma}- k_{\mu\nu }H -\frac{1}{2}\left(K^2-h^2\right)g_{\mu\nu}\;,
\end{equation}
Here $h^2= g^{\mu\nu}k_{\mu\nu}$, $K^{2}=k^{\mu\nu}k_{\mu\nu}$. It follows that $Q_{\mu\nu}$ is conserved in the sense that
\begin{equation}\label{eq:cons}
  Q^{\mu\nu}{}_{;\nu}=0\;.
\end{equation}
The general solution of (\ref{eq:BE2}) for the FLRW geometry was  found  to be
\begin{eqnarray}
&&k_{ij}=\frac{b}{a^2}g_{ij},\;\;i,j=1,2,3, \;\;\;\; k_{44}=\frac{-1}{\dot{a}}\frac{d}{dt}\frac{b}{a}\;, \label{eq:k}
\end{eqnarray}
where we notice that the function $b(t)=k_{11}$ remains an arbitrary function of time. As  a direct consequence of the confinement of the gauge  fields,  Eq. (\ref{eq:BE2}) is   homogeneous,  meaning that
one    component  $k_{11}=b(t)$  remains  arbitrary.
Denoting the Hubble and the extrinsic parameters by $H = \dot{a}/a$ and $B= \dot{b}/b$, respectively, we may write   all components of the extrinsic  geometry in terms  of  $B/H$  as  follows
\begin{eqnarray}
\label{eq:BB}
 &&
 k_{44}=-\frac{b}{a^{2}}(\frac{B}{H}-1)g_{44},\;\;  \\
&&K^{2}=\frac{b^2}{a^4}\left( \frac{B^2}{H^2}-2\frac BH+4\right),
 \;\;\, h=\frac{b}{a^2}(\frac BH+2)\label{eq:hk}\\
&&Q_{ij}= \frac{b^{2}}{a^{4}}\left( 2\frac{B}{H}-1\right)
g_{ij},\; \quad \quad Q_{44} = -\frac{3b^{2}}{a^{4}},
  \label{eq:Qab}\\
&&Q= -(K^2 -h^2) =\frac{6b^{2}}{a^{4}} \frac{B}{H}\;, \label{Q}
 \end{eqnarray}
Next,  by replacing the above results in (\ref{eq:BE1})  and  applying the conservation laws, we obtain the Friedmann equation modified by the presence of the extrinsic curvature, i.e.,
\begin{equation}\label{Friedman}
\left(\frac{\dot{a}}{a}\right)^2+\frac{\kappa}{a^2}=\frac{4}{3}\pi G\rho+\frac{\Lambda_*}{3}+\frac{b^2}{a^4}\;\;.
\end{equation}
When    compared  with the  phenomenological  quintessence phenomenology with constant EoS  we have  found  a  very  close
match  with   the  golden set  of cosmological   data  on the
accelerated expansion  of the universe.

Notice  that   we have  not  used  the  Israel-Lanczos  condition
$ k_{\mu\nu} =\alpha_*(T_{\mu\nu}-  1/3 T g_{\mu\nu})$  as  used  in  \cite{RS}.  If  we  do  so,  in the  case of   the  usual perfect  fluid  matter, then  we obtain  in  Eq. (\ref{Friedman})  a  term  proportional  to  $\rho^2$~\cite{Fried1}. It is possible  to  argue that the above energy-momentum  tensor  $T_{\mu\nu}$  also  include  a  dark energy  component  in  the  energy  density  $\rho$.  However, in this  case  we gain  nothing  because  we  will be  still  in  darkness,  regarding the  nature of  this  energy.  Finally,
as it was  shown  in paper I,  the Israel-Lanczos  condition requires  that  the  four-dimensional  space-time  behaves like a  boundary  brane-world,  with  a mirror  symmetry on it,  which  is  not  compatible  with the regularity  condition for local  and  differentiable  embedding.

Therefore, the  conclusion from paper I is   that the extrinsic  curvature is    a good  candidate  for   the universe accelerator. In the next section  we  start  anew,  with a  mathematical  explanation  on  why  only  gravitation access   the extra  dimensions  using the mentioned   theorem of  Nash on  local  embeddings,  and the   geometric properties of  spin-2  fields defined on space-times.

\section{The Extrinsic Curvature  as  a Dynamical  field}

As it is  well  known,     Riemann geometry  is  determined by the metric  alone,  without appeal to any  external  component.  An  alternative  view, as  used e.g.  in  string theory,  is that of  the  embedded Riemannian   geometry,  which  regards  a  Riemannian  manifold  as embedded into another,   acting as   a  reference  space,  just like  the the Euclidean  3-space acts  as  a reference  to 2-dimensional  surfaces.  However, unlike  the  2-dimensional global embeddings  world-sheets  of  string theory,  the  embedding   of     n-dimensional  manifolds   is   far  more  difficult.  The problem   was   solved in general  form   for  local  embeddings using  differentiable (non-analytic) properties by  John  Nash in 1956.  This  is not  the place   for  a   review  of  Nash's  theorem,  although   its  main properties  have  a clear
application  to gravitational  perturbation theory.

In  short, starting with an  embedded Riemannian  manifold  with    metric $\bar{g}_{\mu\nu}$  and  extrinsic  curvature $\bar{k}_{\mu\nu}$, a new Riemannian geometry with metric $g_{\mu\nu}= \bar{g}_{\mu\nu} + \delta  g_{\mu\nu} $   is  generated,  where
\be
\delta g_{\mu\nu} =-2\bar{k}_{\mu\nu}\delta y\;, \label{eq:York}
\ee
and $\delta y$ is an infinitesimal displacement of the extra dimension $y$. Using this new metric, we obtain a new extrinsic curvature $k_{\mu\nu }$,  and  by   repeating the process a continuous sequence of perturbations may be constructed:
\be
g_{\mu\nu}  =  \bar{g}_{\mu\nu}  +  \delta y \, \bar{k}_{\mu\nu}  +
(\delta y)^2\, \bar{g}^{\rho\sigma}
\bar{k}_{\mu\rho}\bar{k}_{\nu\sigma}\cdots  \label{eq:pertu}
\ee

The embedding  apparently  introduces   fixed background  geometry as opposed to    a  completely intrinsic  and  self-contained  geometry  in  general relativity.  This  can  be  solved by  defining  the  geometry of  the  embedding  space  by  the  Einstein-Hilbert  variational principle,  which  has the  meaning  that   the  embedding  space  has the  smoothest possible  curvature.  This  is  compatible  with   Nash's  theorem  which  requires  a  differentiable  embedding  structure  \cite{QBW}.

Another aspect of  Nash's  theorem is  that  the  extrinsic  curvature  are   the  generator of the  perturbations  of the
gravitational  field  along the  extra  dimensions.  The  symmetric  rank-2  tensor  structure of  the  extrinsic  curvature
lends  the  physical  interpretation  of an    independent   spin-2  field on  the    embedded  space-time.
The study of linear massless spin-2 fields in Minkowski space-time dates back to late 1930s~\cite{FierzPauli}. Some years later, Gupta \cite{Gupta} noted that the Fierz-Pauli equation has a remarkable resemblance with the linear approximation of Einstein's equations for the gravitational field, suggesting that such equation  could be just the linear approximation of a more general, non-linear equation for massless spin-2 fields. In reality, he also found that any spin-2 field in Minkowski space-time must satisfy an equation that has the same formal structure as Einstein's equations. This amounts to saying that, in the same way as Einstein's  equations can be obtained by an infinite sequence of infinitesimal perturbations of the linear gravitational equation, it is possible to obtain  a non-linear equation for any spin-2 field by applying an infinite sequence of infinitesimal perturbations to the Fierz-Pauli equations. The result obtained by  S. Gupta is an Einstein-like system of equations \cite{Fronsdal,Gupta}. In the following, we apply Gupta's equations for the specific case of the extrinsic  curvature of the  FLRW cosmology embedded in a space of 5-dimensions.

The  extrinsic  curvature  $k_{\mu\nu}$  can  be  described as  a  spin-2 field in  an  embedded Riemannian geometry with metric $g_{\mu\nu}$ as  a  solution of   Gupta's  equations.  To write
these  equations  for $k_{\mu\nu}$ we    use an analogy with the derivation of the Riemann  tensor, defining the ``connection" associated with $k_{\mu\nu}$  and then, in analogy  with  the metric, find  the corresponding Riemann tensor, but keeping in mind that the geometry of the embedded space-time was previously   defined by the metric tensor $g_{\mu\nu}$. Let us define  the  tensor
\be
 f_{\mu\nu} = \frac{2}{K}k_{\mu\nu}, \;\; \mbox{and}
\;\;f^{\mu\nu} = \frac{2}{K}k^{\mu\nu}\;,
\label{eq:fmunu}
\ee
so that $f^{\mu\rho}f_{\rho\nu} =\delta^\mu_\nu$. Subsequently, we construct the ``Levi-Civita  connection" associated with $f_{\mu\nu}$, based on the analogy with  the ``metricity condition".  Let us denote by $||$ the covariant derivative with respect to $f_{\mu\nu}$ (while keeping the usual $(;)$ notation for the covariant derivative with respect to $g_{\mu\nu}$), so that $f_{\mu\nu||\rho}=0$. With this condition  we obtain the  ``f-connection"
$$
\Upsilon_{\mu\nu\sigma}=\;\frac{1}{2}\left(\partial_\mu\; f_{\sigma\nu}+ \partial_\nu\;f_{\sigma\mu} -\partial_\sigma\;f_{\mu\nu}\right)
$$
and
$$
\Upsilon_{\mu\nu}{}^{\lambda}= f^{\lambda\sigma}\;\Upsilon_{\mu\nu\sigma}
$$
The ``f-Riemann tensor" associated with this f-connection is
$$
\mathcal{F}_{\nu\alpha\lambda\mu}= \;\partial_{\alpha}\Upsilon_{\mu\lambda\nu}- \;\partial_{\lambda}\Upsilon_{\mu\alpha\nu}+ \Upsilon_{\alpha\sigma\mu}\Upsilon_{\lambda\nu}^{\sigma} -\Upsilon_{\lambda\sigma\mu}\Upsilon_{\alpha\nu}^{\sigma}
$$
and the ``f-Ricci tensor" and the ``f-Ricci scalar", defined with $f_{\mu\nu}$ are, respectively,
$$
{\cal F}_{\mu\nu} =  f^{\alpha\lambda}\mathcal{F}_{\nu\alpha\lambda\mu}
\;\;\mbox{and}\;\;\mathcal{F}=f^{\mu\nu}\mathcal{F}_{\mu\nu}
$$
Finally, write  the  Gupta equations for the $f_{\mu\nu}$ field
\begin{equation}
\label{eq:gupta}
\mathcal{F}_{\mu\nu}-\frac{1}{2}\mathcal{F} f_{\mu\nu} =\;\alpha_f\tau_{\mu\nu}
\end{equation}
where $\tau_{\mu\nu}$ stands for the source of the f-field, with coupling constant $\alpha_f$.  Note that the  above  equation  can  be  derived  from the  action
$$
\delta  \int{\cal F}\sqrt{|f|}dv =0
$$

Note also that, unlike the case of  Einstein's  equations,  here we have not the equivalent  to  the Newtonian weak field limit,  so that
we  cannot   tell  about   the  nature of the source  term  $\tau_{\mu\nu}$. For this reason,  we  start   with  the simplest   Ricci-flat-like equation  for  $f_{\mu\nu}$, i.e.,
\begin{equation}
\label{eq:guptaflat} \mathcal{F}_{\mu\nu}=0\;.
\end{equation}

\section{The Extrinsic   Accelerator}

Although Nash's theorem holds for an arbitrary number of extra dimensions, here we  consider only   one extra dimensions, which,  as  shown  in paper I  is all we need for the local embedding of the FLRW model. Using the definition \rf{k} and (\ref{eq:fmunu}), we obtain the  components of  $f_{\mu\nu}$
\begin{equation}
f_{ij}  =\frac{2}{K} g_{ij},\;\;\; i,j  = 1..3
\end{equation}
and
\begin{equation}
f_{44}  = -\frac{2}{K}\frac{1}{\dot{a}}\frac{d}{dt}{\left(\frac{b}{a}\right)}
\end{equation}
from  which  we  calculate the  components of the  f-connection  $\Upsilon_{\mu\nu\sigma}$,   and of the  the curvature  ${\cal F}_{\mu\nu\rho\sigma}$.  Using the  contractions of the  curvature tensor  with $f_{\mu\nu}$  as  described,  the
equation \rf{guptaflat} is  then  assembled  so that we  may solve Gupta's  equations  to determine  $b(t)$.

For  notational  simplicity  write  $\xi=-k_{44}$. Then   the  components of   (\ref{eq:guptaflat}) are  written as
\begin{equation}
\mathcal{F}_{11}=\frac{1}{4} \frac{-4b^2\xi\dot{K}^2 + 5b\xi\dot{K}\dot{b}K-\dot{b}^2\xi K^2 + 2b^2\xi K \ddot{K} -
2b\ddot{b}\xi K^2 - b^2\dot{K}\dot{\xi}K + bK^2\dot{b}\dot{\xi}}{\xi^2 K^2 b} \label{eq:11}=0
\end{equation}
\begin{equation}
\mathcal{F}_{22}=r^2 \frac{-4b^2\xi\dot{K}^2 + 5b\xi\dot{K}\dot{b}K-\dot{b}^2\xi K^2 + 2b^2\xi K \ddot{K} - 2b\ddot{b}\xi K^2 - b^2\dot{K}\dot{\xi}K + bK^2\dot{b}\dot{\xi}}{4\xi^2 K^2 b}=0
\end{equation}
\begin{equation}
\mathcal{F}_{33}=\sin^2(\theta)\mathcal{F}_{22}=0
\end{equation}
\begin{equation}
\mathcal{F}_{44}=-3/4 \frac{\dot{b}^2\xi K^2 + 2b^2\xi K \ddot{K} -2b\ddot{b}\xi K^2 - b^2\dot{K}\dot{\xi}K + b K^2\dot{b}\dot{\xi} -2b^2\xi \dot{K}^2 + b \xi K \dot{K}\dot{b}}{\xi K^2  b^2} =0 \label{eq:44}
\end{equation}
Note that ${\cal F}_{33}$ is determined from ${\cal F}_{22}$. Therefore,  the only essential equations are   (\ref{eq:11}) and (\ref{eq:44}). By subtracting these equations we obtain $b^2\dot{K}^2 + K^2\dot{b}^2 = 2bK\dot{b}\dot{K}$ or, equivalently,
\be
\left(\frac{\dot{K}}{K}\right)^2 - 2\frac{\dot{b}}{b} \frac{\dot{K}}{K} = -\left(\frac{\dot{b}}{b}\right)^2\;,
\ee
This  equation   has a simple solution which,  for  convenience  we  write as  $K(t)= 2\eta_0 b(t)$, where $2\eta_0$ stands  for the   integration constant. Now, replacing $K$ by its expression in terms of $B$ and $H$  given by (\ref{eq:hk}), we obtain
\begin{equation}\label{eq:EG1}
\frac{B}{H}=1\pm \sqrt{4 \eta_0^2 a^4 - 3}
\end{equation}
As  in paper I,  here  we  use  the conservation   of $Q_{\mu\nu}$ [Eq. (\ref{eq:cons})] providing  the  condition  $2{B}/{H}-1=  \beta_0$, where $\beta_0$ is another integration constant. By subtracting this condition from   \rf{EG1},  we obtain the differentiable equation on $b(t)$ as a function of the expansion parameter $a(t)$, i.e.,
\be
\frac{\dot{b}}{b} =\frac{\dot{a}}{a}( \beta_0  \mp \sqrt{4\eta_0^2 a^4 -3})
\ee
with the general  solution
\begin{equation}
b(t)  =  \alpha_0 a^{\beta_0} e^{\mp \gamma (a)}\;,  \label{eq:b1}
\end{equation}
where  $\alpha_0$ is  another   integration constant  and $\gamma(a)$ is given by
\begin{equation}
\!\!\gamma(a) = 
\! \!\sqrt{4\eta_0^2 a^4 - 3}- \! \sqrt{3}\arctan\left(\frac{\sqrt{3}}{3}\sqrt{4\eta_0^2 a^4 - 3}\right)\;.
\label{eq:gamma}
\end{equation}
Since  $\gamma (z)$  is  real,  we must have (in terms of the redshift $z$)
\begin{equation} \label{eta}
\eta_0^2 \ge \frac{3}{4}(1+z)^4
\end{equation}
where  the equal sign  solution  [$\gamma (z)=0$]  corresponds  to  the  phenomenological fluid model discussed in paper I and $a = 1/(1+z)$. On  the other hand,  the greater  sign provides a more general  solution describing the   acceleration of the universe,  as  we shall see. Replacing this  solution in the  modified  Friedmann's equation (\ref{Friedman})  written  in terms of  the  cosmic  acceleration rate relative to  its  present  value ${{E(a)}}$,  we obtain
\begin{equation} \label{fr}
{{E(z)}}=\frac{\dot{a}}{a} = \left[\Omega_m(1+z)^{3} + \Omega_{\Lambda} + {\Omega_k (1+z)^{2}} + {\Omega_{\rm{ext}} (1+z)^{4-2\beta_0}}e^{\mp \gamma (a)}\right]^{1/2}
\end{equation}
where  $\Omega_m$, $\Omega_{\Lambda}$ and $\Omega_k$ are, respectively, the current values of the matter, cosmological constant and curvature density parameters whereas  $\Omega_{\rm{ext}} = (b_0/H_0)^2$ stands for the density parameter associated with the extrinsic curvature term.

\begin{figure*}
\centerline{
\psfig{figure=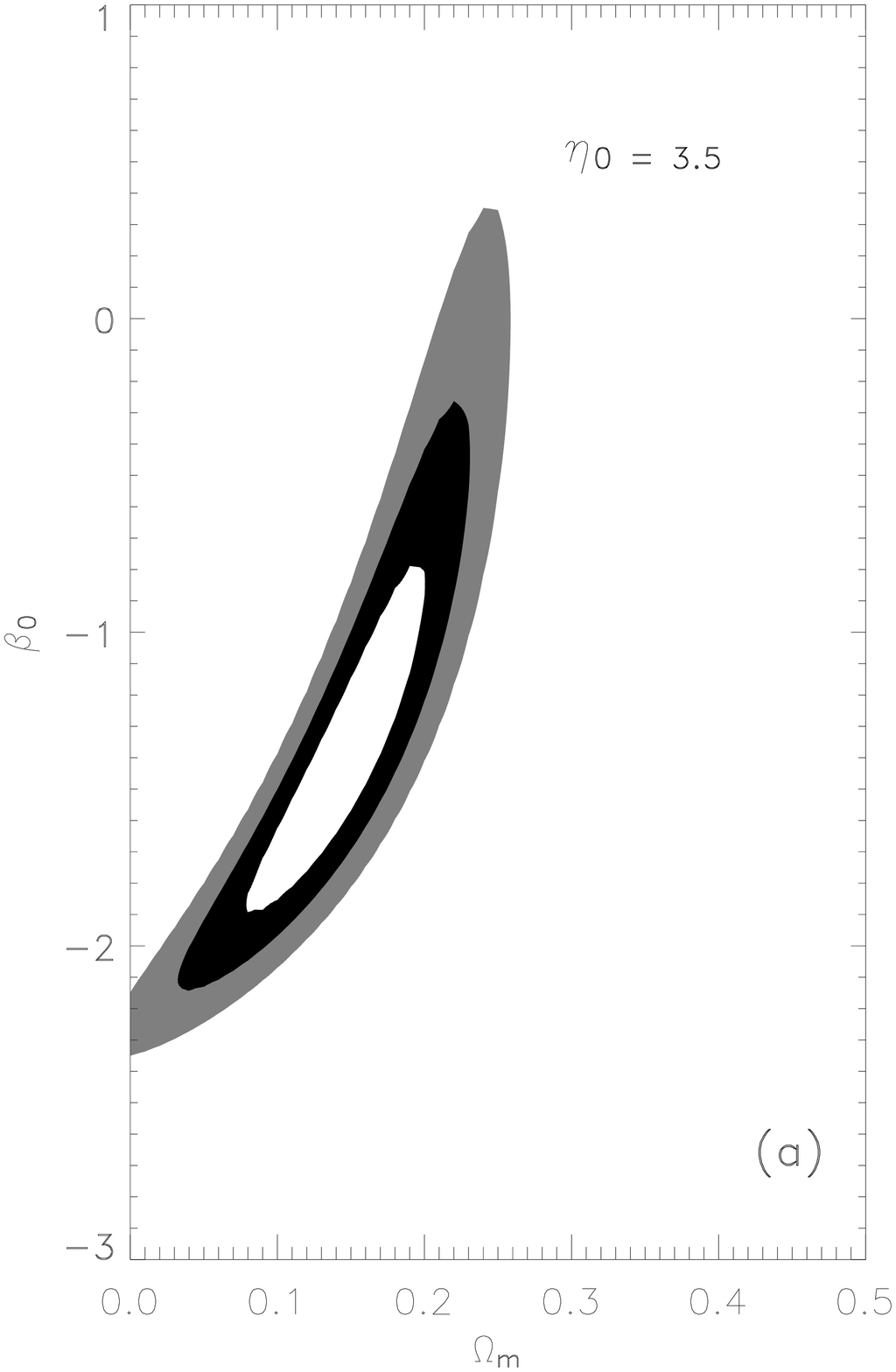,width=2.3truein,height=2.7truein,angle=0}
\psfig{figure=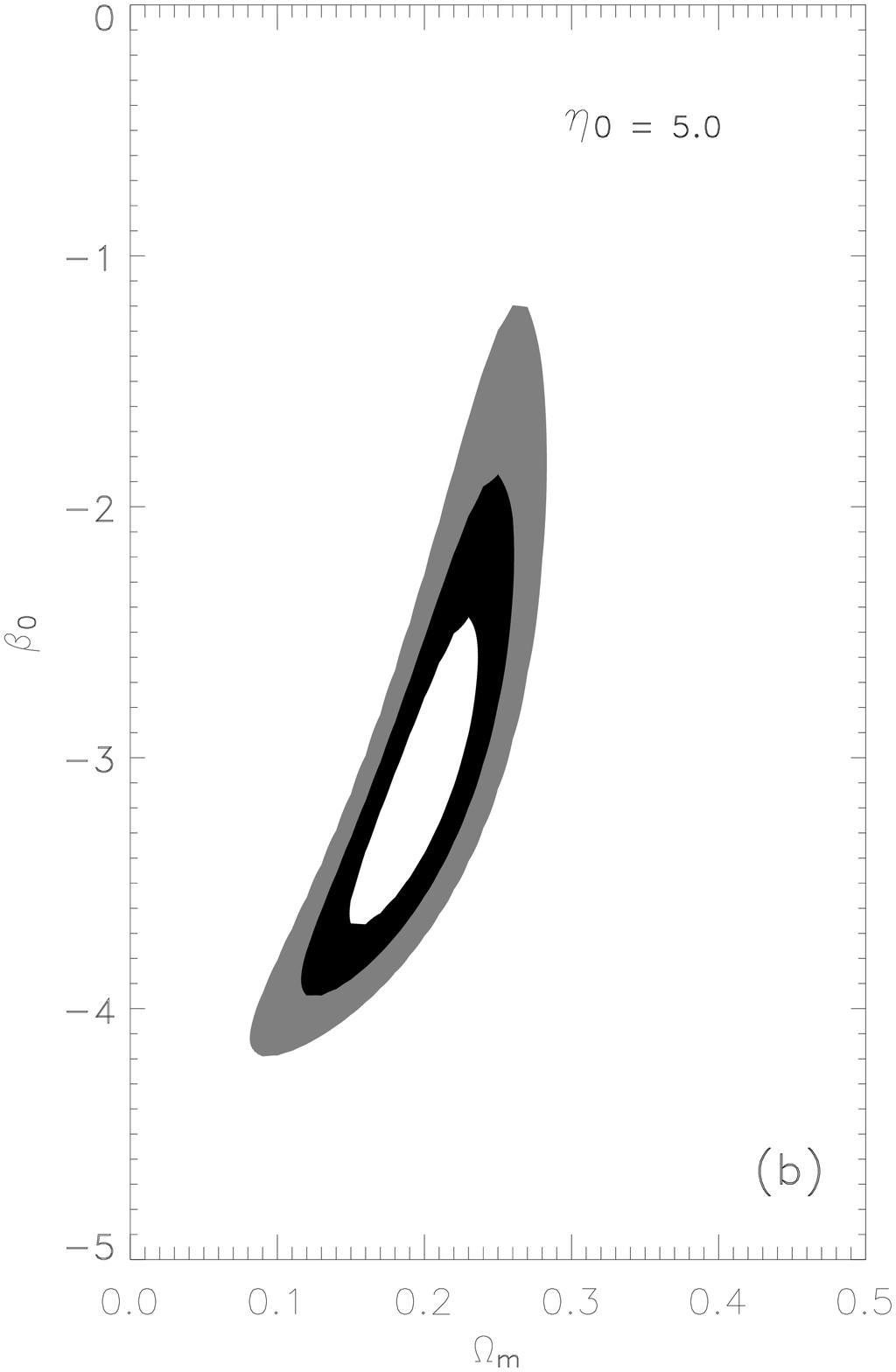,width=2.3truein,height=2.7truein,angle=0}
\psfig{figure=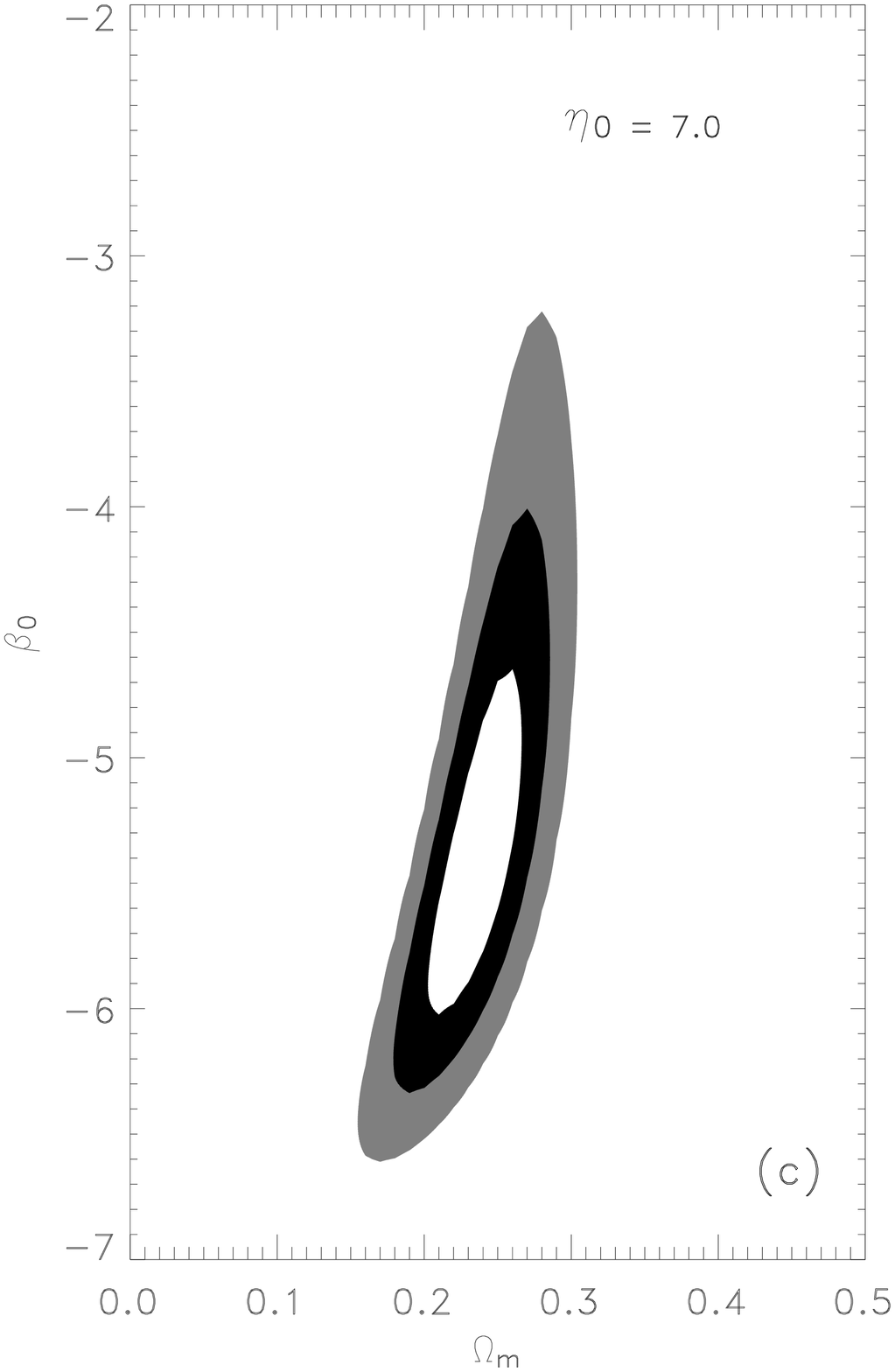,width=2.3truein,height=2.7truein,angle=0}
}
\caption{Contours of $\chi^2$ in the parametric space $\Omega_m - \beta_0$. In all panels, the contours are drawn for $\Delta \chi^2 = 2.30$ and $6.17$. As explained in the text, the value of $\eta_0$ has been fixed at 3.5 (a), 5.0 (b) and 7.0 (c). In particular, we note that for $\eta_0 = 7.0$, the allowed 1$\sigma$ interval for the matter density parameter is very close to that provided by current dynamical estimates, i.e., $\Omega_m \simeq 0.2-0.3$.}
\end{figure*}

In order to study the acceleration phenomenon in these scenarios we use the measured distance-redshift relation of SNe Ia, which provides the most direct probe of the current accelerating expansion. In our analysis we eliminate  the contribution of $\Omega_{\Lambda}$ in Eq. (\ref{fr}) to  emphasize  the  relevance of  the  extrinsic curvature  to the accelerated  expansion of the  universe. Motivated by recent CMB results~\cite{wmap} we also assume $\Omega_k = 0$ and use the normalization condition $\Omega_{\rm{ext}} = (1-\Omega_m)/e^{\gamma (0)}$.

In a flat universe, the dimensionless luminosity-distance is written as
\begin{equation}
d_L(z) H_0= (1+z)\Gamma(z)
\end{equation}
where $\Gamma(z) = \int_0^{z}{dz'/{{E}}(z')}$ and the function $E(z')$ is given by Eq. (\ref{fr}). In our analysis, we use the SNLS collaboration sample of 115 SNe Ia published by Astier {\it et al.} in Ref.~\cite{snls} (for more details on SNe Ia statistical analysis we refer the reader to~\cite{sne,sn1,sn2,sn3,sn4,sn5,sn6} and Refs. therein).

Figure (1a)-(1c) show the results of our statistical analysis. Contours of constant $\Delta \chi^2 = 2.30$, 6.17 and 11.8 are displayed in the $\Omega_m - \beta_0$ space for three different values of $\eta_0$. Since the highest-$z$ SNe Ia in our sample is at $z \simeq 1.01$, we note that the constraint (\ref{eta}) implies $\eta_0 \geq 3.5$, which is the first value considered. The other two values, i.e., $\eta_0 = 5.0$ and $\eta_0 = 7.0$ are taken arbitrarily. At 68.3\% (C.L.), we have found for $\eta_0$ = 3.5, 5.0 and 7.0, respectively,
\[
\beta_0 = -1.45^{+0.30}_{-0.25} \quad \mbox{and} \quad \Omega_m = 0.14 \pm 0.03\;,
\]
\[
\beta_0 = -3.09^{+0.5}_{-0.4} \quad \mbox{and} \quad \Omega_m = 0.20 \pm 0.03\;,
\]
and
\[
\beta_0 = -5.35^{+0.7}_{-0.6} \quad \mbox{and} \quad \Omega_m = 0.24 \pm 0.03\;.
\]
By combining the above results with the normalization condition $\Omega_{\rm{ext}} = (1-\Omega_m)/e^{\gamma (0)}$ obtained from Eq. (\ref{fr}),  we estimate the extrinsic curvature density parameter to lie in the interval $10^{-2} \lesssim \Omega_{\rm{ext}} \lesssim 10^{-6}$.


\section{Final Remarks}

The four-dimensionality  of  space-times is  a  consequence of  the  well established  experimental  structure of  special  relativity,   particle  physics and  quantum  field theory,  using  only the  observables   which interact  with the standard gauge  fields and their  dual properties.  This  has been  described  as  confined quantities,  and it  includes  all  observations  that  are made through  gauge interactions. Any other  observed effects,  usually labeled  as   ''dark", are  known   only    through  their  gravitational consequences.  Therefore, it is   possible that  the  such  ''darkness"  is associated  with  the fact  that  the gravitational  field is not  a gauge  field,  and  consequently it is not necessarily confined. The  only known property of  Riemannian  geometry   describing the perturbations   of  geometry along the extra  dimensions  is  Nash's  theorem  on  local and differentiable perturbative   embedded  Riemannian  manifolds.

In this  paper  we have  described the current cosmic acceleration as a consequence of the extrinsic curvature of the FLRW  universe,  locally embedded in a 5-dimensional   space  defined by the Einstein-Hilbert action. As discussed in Sec. III, Nash's theorem  uses the  extrinsic  curvature  as field  which �provides the
propagation  of the  gravitational  field  along the extra dimensions.  However, as  a  consequence of the  four-dimensional   confinement  of  gauge  fields  and  ordinary matter,  the  extrinsic  curvature  is not  completely determined by  the embedding  equations.  Therefore, in order to   complement  the number of  required equations,   we  have  noted that the  extrinsic  curvature   is a rank-2 symmetric tensor, which   corresponds  to  a spin-2 field defined on the embedded space-time. As it was demonstrated by Gupta,  any spin-2  field  satisfies an Einstein-like equation. After the due adaption to  an  embedded space-time,  we have constructed the Gupta  equations for the extrinsic curvature of  the FLWR geometry and studied the behavior  of its  solution  at  the current stage of the cosmic evolution.  We have also tested the observational viability of these scenarios by confronting their theoretical predictions for an accelerating universe with current SNe Ia data. We have shown that a very small contribution of $\Omega_{\rm{ext}}$ ($\sim 10^{-2} - 10^{-6}$) is enough to provide a possible explanation for the current observed  accelerated  expansion of the Universe.

\end{document}